\documentclass[a4paper,twoside]{article}

\usepackage{epsfig}
\usepackage{subfigure}
\usepackage{calc}
\usepackage{amssymb}
\usepackage{amstext}
\usepackage{amsmath}
\usepackage{amsthm}
\usepackage{multicol}
\usepackage{pslatex}
\usepackage{apalike}
\usepackage{SCITEPRESS}     

\usepackage{booktabs}
\usepackage{url}

\begin{document}

\title{About being the Tortoise or the Hare? \subtitle{A Position Paper on Making Cloud Applications too Fast and Furious for Attackers} }

\author{\authorname{Nane Kratzke}
\affiliation{L\"ubeck University of Applied Sciences\\
	Center of Excellence for Communication, Systems, and Applications (CoSA)\\
	M\"onkhofer Weg 239, 23562 L\"ubeck, Germany}
\email{nane.kratzke@fh-luebeck.de}
}


\abstract{ 
	Cloud applications expose -- beside service endpoints -- also potential or actual vulnerabilities.
	And attackers have several advantages on their side. They can select the weapons, the point of time and the point of attack.
	Very often cloud application security engineering efforts focus to harden the fortress walls but seldom assume that attacks may be successful. So, cloud applications rely on their defensive walls but seldom attack intruders actively. Biological systems are different. They accept that defensive ``walls" can be breached at several layers and therefore make use of an active and adaptive defense system to attack potential intruders - an immune system. 
	This position paper proposes such an immune system inspired approach to ensure that even undetected intruders can be purged out of cloud applications. This makes it much harder for intruders to maintain a presence on victim systems. Evaluation experiments with popular cloud service infrastructures (Amazon Web Services, Google Compute Engine, Azure and OpenStack) showed that this could minimize the undetected acting period of intruders down to minutes.
}

\onecolumn \maketitle \normalsize \vfill

\section{\uppercase{Introduction}}

\begin{table}[b]
	\caption{\textbf{Some popular open source elastic platforms}}
	\label{tab:platforms} 
	\centering
	\scriptsize
	\begin{tabular}{lll}
		\toprule
		\textbf{Platform} & \textbf{Contributors} & \textbf{URL}\\
		\midrule
		Kubernetes & Cloud Native Found. &  \url{http://kubernetes.io}\\
		Swarm         & Docker & \url{https://docker.io} \\
		Mesos         & Apache & \url{http://mesos.apache.org/} \\
		Nomad     & Hashicorp & \url{https://nomadproject.io/} \\
		\bottomrule
	\end{tabular}
\end{table}

\noindent ``The Tortoise and the Hare" is one of Aesop's most famous fables where ingenuity and trickery are employed by the tortoise to overcome a stronger opponent -- the hare. Regarding this paper and according to this fable, the hare is an attacker and the tortoise is an operation entity responsible to protect a cloud system against security breaches. Zero-day exploits make this game an unfair game. How to protect a cloud system against threats that are unknown to the operator? But, when the game itself is unfair, should not the system operation entity be unfair as well?  That is basically what this position paper is about. How to build ``unfair" cloud systems that permanently jangle attackers nerves.

\begin{figure*}
	\centering%
	\includegraphics[width=0.7\textwidth]{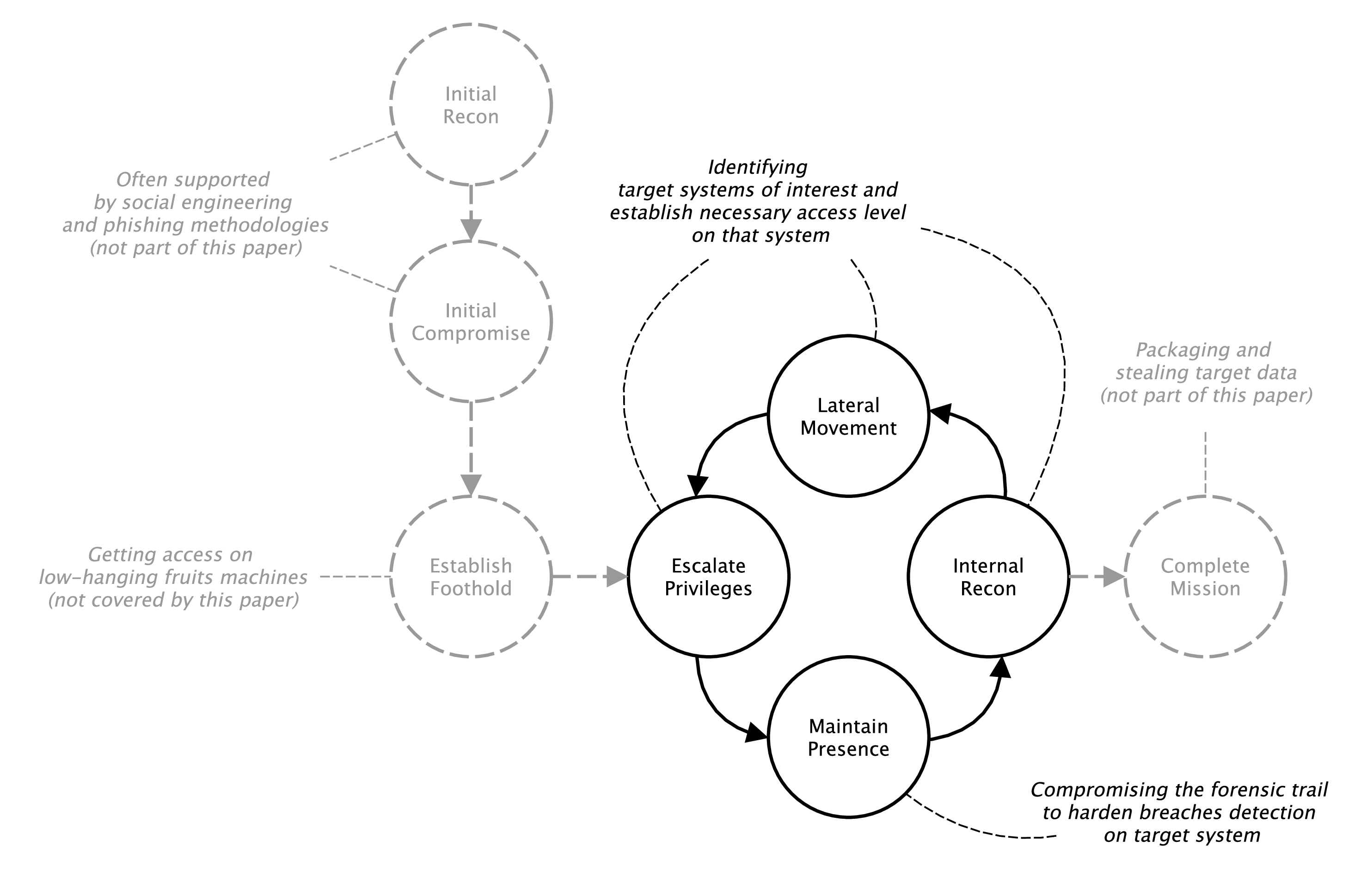}
	
	\caption{\textbf{The cyber attack life cycle model.} \textit{Adapted from the cyber attack lifecycle used by the M-Trends reports, see Table \ref{tab:dwells}}\label{fig:cyber-attack}.}
\end{figure*}

Cloud computing enables a variety of innovative IT-enabled business and service models and many research studies and programs focus to develop systems in a responsible way to ensure the security and privacy of users. But compliance with standards, audits and checklists, does not automatically equal security \cite{DW2014} and there is a fundamental issue remaining. 
Zero-day vulnerabilities are computer-software vulnerabilities that are unknown to those who would be interested in mitigating the vulnerability (including the entity responsible to operate a cloud application). Until a vulnerability is mitigated, hackers can exploit it to adversely affect computer programs, data, additional computers or a network. For zero-day exploits, the probability that vulnerabilities are patched is zero, so the exploit should always succeed. Therefore, zero-day attacks are a severe threat and we have to draw a scary conclusion: \textbf{In principle attackers can establish footholds in our systems whenever they want.}

Recent research \cite{Kra2017a,Kra2018} made successfully use of elastic container platforms (see Table \ref{tab:platforms}) and their ``designed for failure" capabilities to realize transferability of cloud-native applications at runtime. By transferability, the conducted research means that a cloud-native application can be moved from one IaaS provider infrastructure to another without any downtime. These platforms are more and more used as distributed and elastic runtime environments for cloud-native applications \cite{KQ2017a} and can be understood as a kind of cloud infrastructure unifying middleware \cite{KP2016}. It should be possible to make use of the same features to immunize cloud applications simply by moving an application within the same provider infrastructure. To move anything from A to A makes no sense at first glance. However, let us be paranoid and aware that with some probability and at a given time, an attacker will be successful and compromise at least one virtual machine \cite{BD2012}. In these cases, a transfer from A to A would be an efficient counter measure -- because the intruder immediately loses any hijacked machine that is moved. To understand that, the reader must know that our approach does not effectively move a machine, it regenerates it. To move a machine means to launch a compensating machine unknown to the intruder and to terminate the former (hi-jacked) machine. Whenever an application is moved all of its virtual machines are regenerated. And this would effectively eliminate undetected hi-jacked machines.  

The biological analogy of this strategy is called \textit{``cell-regeneration"} and the attack on ill cells is coordinated by an immune system. This paper describes first ideas for such a kind of immune system following this \textbf{outline}. To provide some context for the reader, Section \ref{sec:lifecycle-cyber-attacks} will explain the general life-cycle of a cyber attack. It is assumed that every system can be penetrated due to zero-day exploits. Section \ref{sec:transferable-systems} will summarize some of our recent research to explain how such immune systems could be built. Section \ref{sec:evaluation} shows some evaluation results measured from transferability experiments. These numbers are used to estimate possible regeneration intervals for systems of different sizes and to compare them with median dwell times reported by security companies over the last seven years (see Table \ref{tab:dwells}). The advantages and limitations of this proposal are related to other work in Section \ref{sec:related_work}. Finally, this proposal is discussed from a more critical point view in Section \ref{sec:critical-discussion} to derive future research challenges in Section \ref{sec:conclusion}.

\section{\uppercase{Cyber Attack Life Cycle}}
\label{sec:lifecycle-cyber-attacks}

\noindent Figure \ref{fig:cyber-attack} shows the cyber attack life cycle model which is used by the M-Trends reports\footnote{\url{http://bit.ly/2m7UAYb} (visited 9th Nov. 2017)}
to report developments in cyber attacks over the years. According to this model, an attacker passes through different stages to complete a cyber attack mission. It starts with initial reconnaissance and compromising of access means. These steps are very often supported by social engineering methodologies \cite{KHH+2015}  and phishing attacks \cite{GSK2016}. The goal is to establish a foothold near the system of interest.  All these steps are not covered by this paper, because technical solutions are not able to harden the weakest point in security -- the human being.  
The following steps of this model are more interesting for this paper. According to the life cycle model the attacker's goal is to escalate privileges to get access to the target system. Because this leaves trails on the system which could reveal a security breach, the attacker is motivated to compromise this forensic trail. According to security reports attackers make more and more use of counter-forensic measures
to hide their presence and impair
investigations. These reports refer to
batch scripts used to clear event
logs and securely delete arbitrary files.
The technique is simple, but the intruders'
knowledge of forensic artifacts
demonstrate increased sophistication, as well as
their intent to persist in the environment. 
With a barely detectable foothold, the internal reconnaissance of the victim's network is carried out to allow the lateral movement to the target system. This is a complex and lengthy process and may even take weeks. So, infiltrated machines have worth for attackers and tend to be used for as long as possible. Table \ref{tab:dwells} shows how astonishingly many days on average an intruder has access to a victim system. So, basically there is the requirement, that \textbf{an undetected attacker should lose access to compromised nodes of a system as fast as possible.} But how?

\begin{table}[t]
	\caption{\textbf{Undetected days on victim systems} \textit{reported by M-Trends. External and internal discovery data is reported since 2015. No data could be found for 2011.}} 
	\label{tab:dwells} 
	\centering
	\scriptsize
	\begin{tabular}{lccr}
		\toprule
		\textbf{Year} & \textbf{External notification} & \textbf{Internal discovery} & \textbf{Median}\\
		\midrule
		2010 & - & - & 416 \\
		2011 & - & - & ? \\
		2012 & - &  - & 243 \\
		2013  & - & - & 229 \\
		2014  & - & - & 205 \\
		2015  & 320 & 56 & 146 \\
		2016  & 107 & 80 & 99 \\
		\bottomrule
	\end{tabular}
\end{table}
\section{\uppercase{Regenerate-able Cloud Applications}}
\label{sec:transferable-systems}

\noindent Our recent research dealt mainly with vendor lock-in and the question how to design cloud-native applications that are transferable between different cloud service providers. One aspect that can be learned from this is that there is no common understanding of what a cloud-native application really is. A kind of software that is \textit{``intentionally designed for the cloud"} is an often heard but vacuous phrase. However, noteworthy similarities exist between various view points on \textit{cloud-native applications} (CNA) \cite{KQ2017a}. A common approach is to define maturity levels in order to categorize different kinds of cloud applications (see Table \ref{tab:camm}).
\cite{FLR+2014} proposed the IDEAL model for CNAs. A CNA should strive for an \textbf{\underline{i}solated state}, is \textbf{\underline{d}istributed}, provides \textbf{\underline{e}lasticity} in a horizontal scaling way, and should be operated on \textbf{\underline{a}utomated deployment machinery}. Finally, its components should be \textbf{\underline{l}oosely coupled}.

\begin{figure*}
	\begin{center}
		\includegraphics[width=0.75\textwidth]{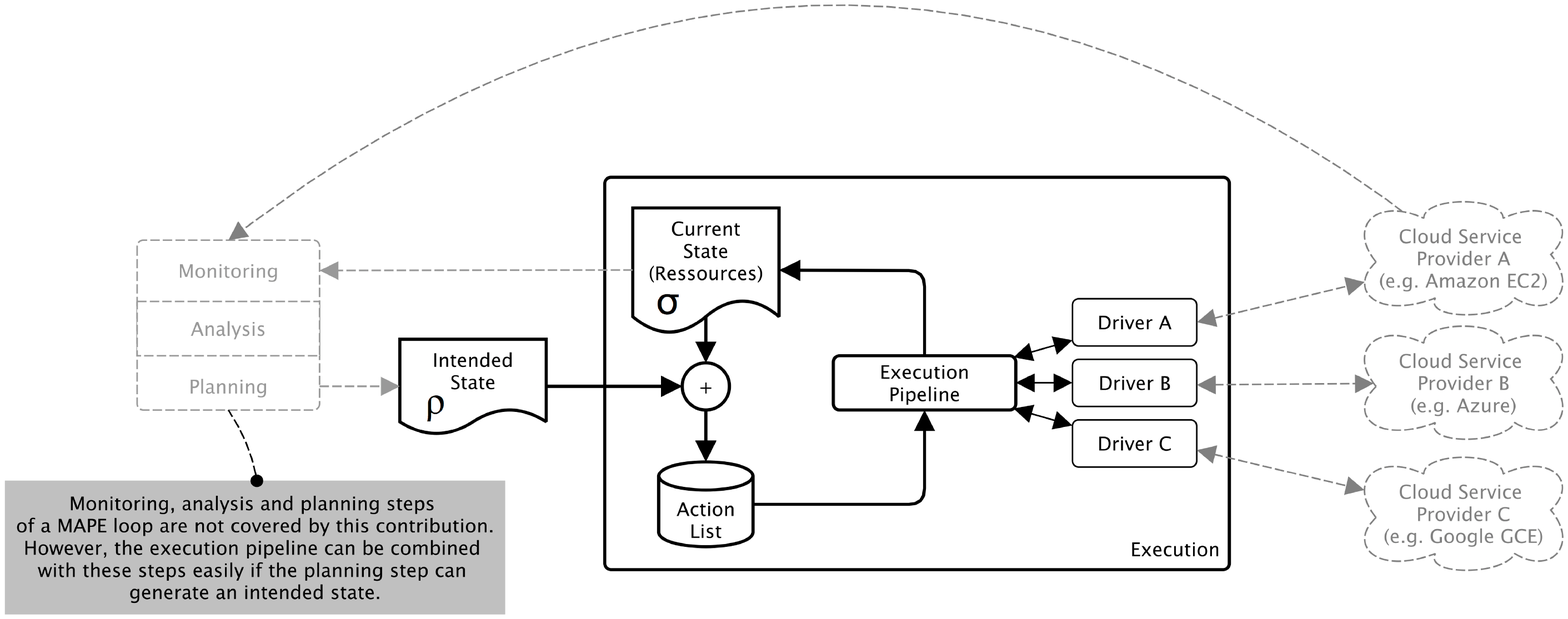}
	\end{center}
	\caption{\textbf{The control theory inspired execution control loop} \textit{compares the intended state $\rho$ of an elastic container platform with the current state $\sigma$ and derives necessary scaling actions. These actions are processed by the execution pipeline explained in Figure \ref{fig:pipeline}. So, platforms can be  operated elastically in a set of synchronized IaaS infrastructures. Explained in details by \cite{Kra2017a}.}}
	\label{fig:executionloop}		
\end{figure*}

\begin{figure*}
	\centering
	\includegraphics[width=0.75\textwidth]{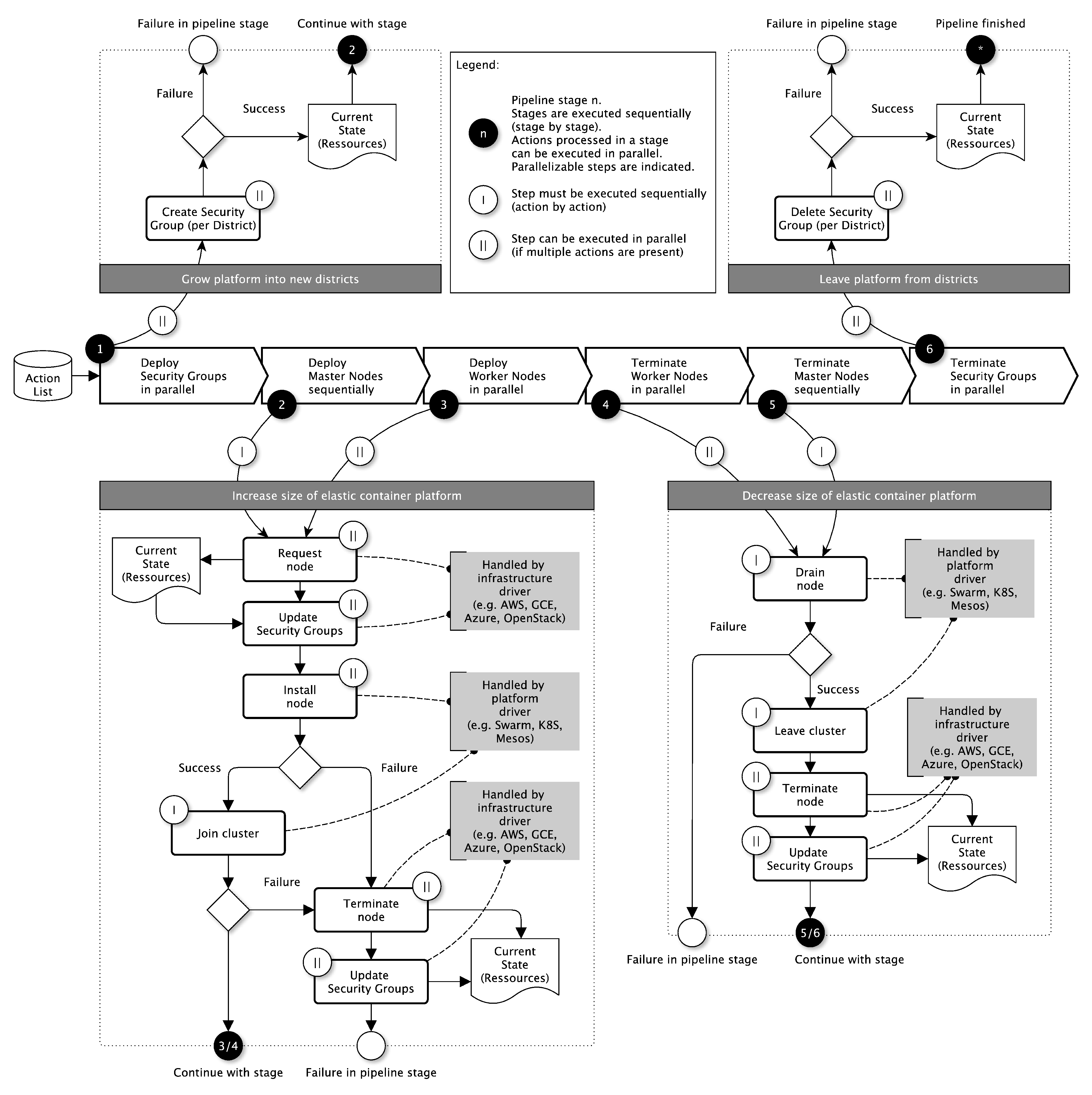}
	\caption{\textbf{The execution pipeline} \textit{processes necessary actions to transfer the current state $\sigma$ into the intended state $\rho$. See \cite{Kra2018} for more details.}}
	\label{fig:pipeline}
\end{figure*}

\begin{table}[b]
	\caption{\textbf{Cloud Application Maturity Model}, adapted from \textit{OPEN DATA CENTER ALLIANCE Best Practices \cite{ODCA2014}}}
	\label{tab:camm} 
	\centering
	\scriptsize
	\begin{tabular}{ccl}
		\toprule
		\textbf{Level} & \textbf{Maturity} & \textbf{Criteria}\\
		\midrule
		3 & Cloud     & - Transferable across infrastructure providers at \\
		& native    &   \enspace runtime and without interruption of service.\\
		&           & - Automatically scale out/in based on stimuli.\\
		\midrule
		2 & Cloud     & - State is isolated in a minimum of services. \\
		& resilient & - Unaffected by dependent service failures. \\
		&           & - Infrastructure agnostic.\\
		\midrule
		1 & Cloud     & - Composed of loosely coupled services. \\
		& friendly  & - Services are discoverable by name. \\
		&           & - Components are designed to cloud patterns. \\
		&           & - Compute and storage are separated. \\
		\midrule
		0 & Cloud     & - Operated on virtualized infrastructure. \\
		& ready     & - Instantiateable from image or script. \\
		\bottomrule
	\end{tabular}
\end{table}

\cite{BHJ2015} stress that these properties are addressed by cloud-specific architecture and infrastructure approaches like \textbf{Microservices} \cite{Newman2015}, \textbf{API-based collaboration}, adaption of \textbf{cloud-focused patterns} \cite{FLR+2014}, and \textbf{self-service elastic platforms} that are used to deploy and operate these microservices via self-contained deployment units (containers). Table \ref{tab:platforms} lists some of these platforms that provide additional operational capabilities on top of IaaS infrastructures like automated and on-demand scaling of application instances, application health management, dynamic routing and load balancing as well as aggregation of logs and metrics \cite{KQ2017a}.

If the reader understands and accepts the commonality that cloud-native applications are operated (more and more often) on elastic -- often container-based -- platforms, it is an obvious idea to delegate the responsibility to immunize cloud applications to these platforms. Recent research showed that the operation of these elastic container platforms and the design of applications running on-top of them should be handled as two different engineering problems. This often solves several issues in modern cloud-native application engineering \cite{Kra2018}. And that is not just true for the transferability problem but might be an option to tackle zero-day exploits. These kind of platforms could be an essential part of the immune system of modern cloud-native applications.

Furthermore, \textbf{self-service elastic platforms} are really ``bulletproofed" \cite{Stine2015}. \textit{Apache Mesos} \cite{mesos} has been successfully operated for years by companies like Twitter or Netflix to consolidate hundreds of thousands of compute nodes. Elastic container platforms are \textbf{designed for failure} and provide self-healing capabilities via auto-placement, auto-restart, auto-replication and auto-scaling features. They will identify lost containers (for whatever reasons, e.g. process failure or node unavailability) and will restart containers and place them on remaining nodes. These features are absolutely necessary to operate large-scale distributed systems in a resilient way. However, the same features can be used intentionally to \textbf{purge ``compromised nodes"}.

\cite{Kra2017a} demonstrated a software prototype that provides the control process shown in Figure \ref{fig:executionloop} and Figure \ref{fig:pipeline}. This process relies on an \textit{intended state} $\rho$ and a \textit{current state} $\sigma$ of a container cluster. If the intended state differs from the current state ($\rho\not = \sigma$), necessary adaption actions are deduced (creation and attachment/detachment of nodes, creation and termination of security groups) and processed by an execution pipeline fully automatically (see Figure \ref{fig:pipeline}) to reach the \textit{intended state} $\rho$. With this kind of control process, a cluster can be simply resized by changing the intended amount of nodes in the cluster. If the cluster is shrinking and nodes have to be terminated, affected containers of running applications will be rescheduled to other available nodes.

The downside of this approach is, that this will only work for Level 2 (cloud resilient) or Level 3 (cloud native) applications (see Table \ref{tab:camm}) which by design, can tolerate dependent service failures (due to node failures and container rescheduling). However, for that kind of Level 2 or Level 3 application, we can use the same control process to regenerate nodes of the container cluster. The reader shall consider a cluster with $\sigma=N$ nodes. If we want to regenerate one node, we change the intended state to $\rho=N+1$ nodes which will add one new node to the cluster ($\sigma'=N+1$). And in a second step, we will decrease the intended size of the cluster to $\rho'=N$ again, which has the effect that one node of the cluster is terminated ($\sigma''=N$). So, a node is regenerated simply by adding one node and deleting one node. We could even regenerate the complete cluster by changing the cluster size in the following way: $\sigma = N \mapsto  \sigma' = 2N \mapsto \sigma'' = N$. But, this would consume much more resources because the cluster would double its size for a limited amount of time. A more resource efficient way would be to regenerate the cluster in $N$ steps: $\sigma = N \mapsto  \sigma' = N+1 \mapsto \sigma'' = N \mapsto ... \mapsto \sigma^{2N-1}=N+1 \mapsto \sigma^{2N}=N$.   This should make the general idea clear. The reader is referred to \cite{Kra2018} for more details, especially if the reader is interested in the multi-cloud capabilities, that are not covered by this paper due to page limitations.

Whenever such a regeneration is triggered, all -- even undetected -- hijacked machines would be terminated and replaced by other machines, but the applications would be unaffected. For an attacker, this means losing their foothold in the system completely. Imagine this would be done once a day or even more frequently?
\section{\uppercase{Evaluation results}}
\label{sec:evaluation}

\begin{table}[b]
	\caption{\textbf{Used machine types and regions for evaluation}}
	\label{tab:machine-types} 
	\centering
	\scriptsize
	\begin{tabular}{llll}
		\toprule
		\textbf{Provider} & \textbf{Region} &\textbf{Master type} & \textbf{Worker type}\\
		\midrule
		AWS & eu-west-1 & m4.xlarge & m4.large \\
		GCE & europe-west1 & n1-standard-4 & n1-standard-2 \\
		Azure & europewest & Standard\_A3 & Standard\_A2 \\
		OS & \textit{own datacenter} & m1.large & m1.medium \\
		\bottomrule
	\end{tabular}
\end{table}

\begin{table}[b]
	\caption{\textbf{Durations to regenerate a node (median values)}}
	\label{tab:regeneration-times} 
	\centering
	\scriptsize
	\begin{tabular}{lrrrrr}
		\toprule
		\textbf{Provider} & \textbf{Creation} &\textbf{Secgroup} & \textbf{Joining} & \textbf{Term.} & \textbf{\underline{Total}}\\
		\midrule
		AWS & 70 s & 1 s & 7 s & 2 s & \textbf{81 s} \\
		GCE & 100 s & 8 s & 9 s & 50 s & \textbf{175 s} \\
		Azure & 380 s & 17 s & 7 s & 180 s & \textbf{600 s} \\
		OS  & 110 s & 2 s & 7 s & 5 s & \textbf{126 s} \\
		\bottomrule
	\end{tabular}
\end{table}

\noindent The execution pipeline presented in Figure \ref{fig:pipeline} was evaluated by operating and transferring two elastic platforms (\textit{Swarm Mode of Docker 17.06} and \textit{Kubernetes 1.7}). The platforms operated a reference ``sock-shop" application being one of the most complete reference applications for microservices architecture research \cite{AMP+2017}. Table \ref{tab:machine-types} lists  the machine types that show a high similarity across different providers \cite{KQ2015}.

\begin{figure}[t]
	\centering
	\includegraphics[width=\columnwidth]{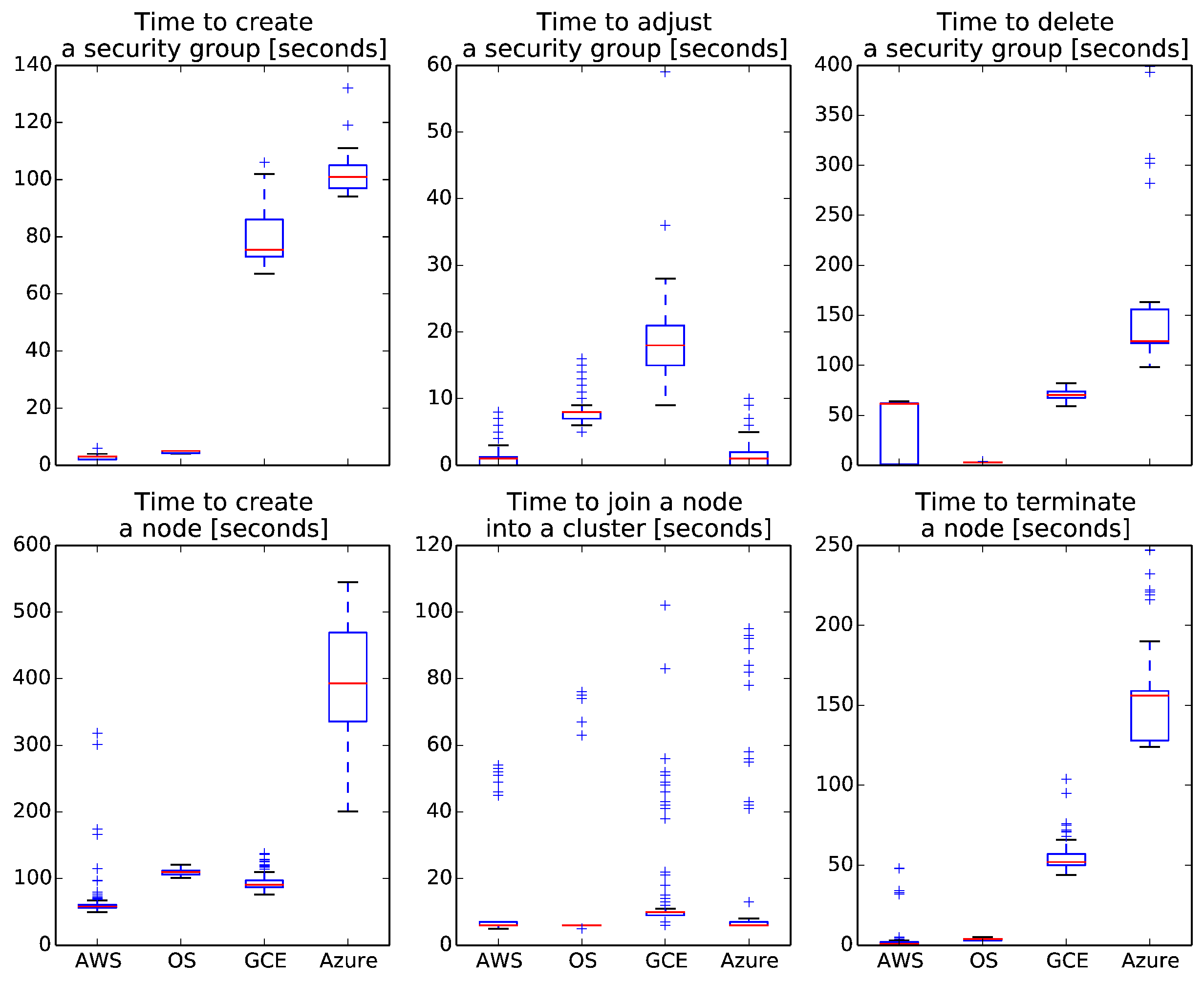}
	\caption{\textbf{Infrastructure specific runtimes of IaaS operations.} \textit{Taken from \cite{Kra2018}.}}
	\label{myfigone}
\end{figure}

The evaluation of \cite{Kra2018} demonstrated that most time is spent on the IaaS level (creation and termination of nodes and security groups) and not on the elastic platform level (joining, draining nodes). The measured differences on infrastructures provided by different providers is shown in Figure \ref{myfigone}. For the current use case the reader can ignore the times to create and delete a security group (because that is a one time action). However, there will be many node creations and terminations. According to our execution pipeline shown in Figure \ref{fig:pipeline}, a node creation ($\sigma = N \mapsto \sigma'=N+1$) involves the durations to \textbf{create a node} (request of the virtual machine including all installation and configuration steps), to \textbf{adjust security groups} the cluster is operated in and to \textbf{join the new node} into the cluster. The shutdown of a node ($\sigma=N \mapsto \sigma'=N-1$) involves the \textbf{termination of the node} (this includes the platform draining and deregistering of the node and the request to terminate the virtual machine) and the necessary \textbf{adjustment of the security group}. So, for a complete regeneration of a node ($\sigma=N \mapsto \sigma'=N+1 \mapsto \sigma''=N$) we have to add these runtimes. Table \ref{tab:regeneration-times} lists these values per infrastructure.

Even on the ``slowest" infrastructure, a node can be regenerated in about 10 minutes. In other words, one can regenerate six nodes every hour or up to 144 nodes a day or a cluster of 432 nodes every 72h (which is the reporting time requested by the EU General Data Protection Regulation). If the reader compares a 72h regeneration time of a more than 400 node cluster (most systems are not so large) with the median value of 99 days that attackers were present on a victim system in 2016 (see Table \ref{tab:dwells}) the benefit of the proposed approach should become obvious.
\section{\uppercase{Related Work}}
\label{sec:related_work}

\noindent To the best of the author's knowledge, there are currently no approaches making intentional use of virtual machine regeneration for security purposes. However, the proposed approach is derived from multi-cloud scenarios and their increased requirements on security. And there are several promising approaches dealing with multi-cloud scenarios. So, all of them could show comparable opportunities. But often, these approaches come along with a lot of inner complexity. A container based approach seems to handle this kind of complexity better. There are some good survey papers on this \cite{Barker2015,Petcu2014,Toosi2014,Grozev2014}.

To secure the forensic trail is essential  for anomaly detection approaches in log data \cite{FLW+2009,WSF+2017}. Therefore \cite{DW2016b,DW2016a} propose to use an immutable database for this purpose, which they suggested to be kept in a remote location from the main cloud system.
Further research deals with append-only data structures on untrusted servers \cite{PPP+2015}. Other approaches propose building a secure and reliable file synchronization service using multiple cloud synchronization services as untrusted storage providers \cite{HST+2015}. Further approaches focus on the integrity of logs and ensure their integrity by hash-chain schemes and proofs of past logs published periodically by the cloud providers \cite{ZDH2016}. The question remains, whether these approaches are scalable enough to provide robust logging means for the forensic trail of up to thousands of nodes. Messaging solutions like Kafka \cite{WKS+2015} or logging stacks like the ELK-Stack are bullet-proofed technologies for consolidating logs but assume to be operated in a trusted environment which often ends in very complicated kind of double logging architectures \cite{Kra2018b}.
\section{\uppercase{Critical Discussion}}
\label{sec:critical-discussion}

\noindent The idea of using an immune system like approach to remove undetected intruders in virtual machines seems to a lot of experts intriguing. But state of the art is, that this is not done. And there might be reasons for that and open questions the reader should consider.

Several reviewers remarked that the proposal can be compared with the approach to  restart periodically virtual machines  that  have memory leak issues. This has nothing to do with security concerns, and could be applied to traditional (non-cloud) systems as well. So, the approach may have even a broader focus than presented (which is not a bad thing).

Another question is  how to detect ``infected" nodes? The presented approach selects nodes simply at random. This will hit every node at some time. The same could be done using a round-robin approach but a round-robin strategy  would be better predictable for an attacker. However, both strategies will create a lot of unnecessary regenerations and that leaves obviously room for improvements. It seems obvious to search for solutions like presented by \cite{FLW+2009,WSF+2017} to provide some  ``intelligence" for the identification of  ``suspicious" nodes. This would limit regenerations to likely ``infected" nodes. In all cases it is essential for anomaly detection approaches  to secure the forensic trail \cite{DW2016b,DW2016a}.

Furthermore, to regenerate nodes periodically or even randomly is likely nontrivial in practice and  depends on the state management requirements for the affected nodes. Therefore, this paper proposes the approach only as a promising solution for Level 2 or 3 cloud applications (see Table \ref{tab:camm}) that are operated on elastic container platforms. That kind of applications have eligible state management characteristics. But, this is obviously a limitation.

One could be further concerned about exploits that are adaptable to bio-inspired systems. Stealthy resident worms dating back to the old PC era would be an example. This might be especially true for the often encountered case of not entirely stateless services, when data-as-code dependencies or code-injection vulnerabilities exist. Furthermore, attackers could shift their focus to the platform itself in order to disable the regeneration mechanism as a first step. On the other hand, this could be easily detected -- but there could exist more sophisticated attacks.

Finally, there is obviously room and need for a much more detailed evaluation. The effectiveness of this approach needs a large scale and real world evaluation with more complex  cloud native applications using multiple coordinated virtual machines. This is up for ongoing research and should be kept in mind.
\section{\uppercase{Conclusion}}
\label{sec:conclusion}

\noindent There is still no such thing as an impenetrable system. Once attackers successfully breach a system, there is little to prevent them from doing arbitrary harm -- but we can reduce the available time for the intruder to do this. The presented approach evolved mainly from transferability research questions for cloud-native applications. But it can be the  foundation for an ``immune system" inspired approach to tackle zero-day exploits. The main intent is simply to massively reduce the time for an attacker acting undetected. Therefore, this paper proposed to regenerate virtual machines (the cells of an IT-system) with a much higher frequency than usual to purge even undetected intruders. Evaluations on infrastructures provided by AWS, GCE, Azure and OpenStack showed that a virtual machine can be regenerated between two minutes (AWS) and 10 minutes (Azure). The reader should compare these times with recent cyber security reports. In 2016 an attacker was undetected on a victim system for about 100 days. The presented approach means for intruders that their undetected time on victim systems is not measured in months or days anymore, it would be measured in minutes.

Such a biology inspired immune system solution is charming but may also involve downsides. To regenerate too many nodes at the same time would let the system run ``hot". The reader might know this health state from own experiences as fever. And if the immune system attacks to many unaffected (healthy) nodes again and again, this could be even called an auto-immune disease.  Both states are not the best operation modes of immune systems. Although the presented approach can limit available time for an attack substantially, we should consider that even in a very short amount of time an attacker could delete (parts) of the cloud forensic trail. This could limit the effectiveness of  ``suspect node" detection mechanisms. To use external and trusted append-only logging systems seems somehow obvious. However, existing solutions rely very often on trusted environments. 

So, further research should investigate how ``regenerating" platforms and append-only logging systems can be operated on untrusted environments without fostering unwanted and non-preferable effects known from the human immune system like fever or even auto-immune diseases. The critical discussion in Section \ref{sec:critical-discussion} showed that there is need for additional evaluation and room for more in-depth research. However, several reviewers remarked independently that the basic idea is so ``intriguing",  that it should be considered more consequently.

\section*{\uppercase{Acknowledgements}}

\noindent This research is partly funded by the Cloud TRANSIT project (13FH021PX4, German Federal Ministry of  Education and Research). I would like to thank Bob Duncan from the University of Aberdeen and all the anonymous reviewers for their inspiring thoughts on cloud security challenges.
 
\bibliographystyle{apalike}
{\small
\bibliography{references}}

\vfill
\end{document}